\documentclass[twocolumn,groupedaddress,amsmath,amssymb]{revtex4}

\usepackage{graphicx}
\usepackage{dcolumn}
\usepackage{bm}% bold math
\usepackage{amsmath,amssymb}

\begin{document}

\title{X-ray magnetic circular dichroism and photoemission 
study of the 
diluted ferromagnetic semiconductor Zn$_{1-x}$Cr$_x$Te}

\author{Y.~Ishida}
\affiliation{Department of Physics, 
University of Tokyo, Bunkyo-ku, Tokyo 113-0033, Japan
}%

\author{M.~Kobayashi}
\affiliation{Department of Physics, 
University of Tokyo, Bunkyo-ku, Tokyo 113-0033, Japan
}%

\author{J.I.~Hwang}
\affiliation{Department of Physics, 
University of Tokyo, Bunkyo-ku, Tokyo 113-0033, Japan
}%

\author{Y.~Takeda}
\affiliation{Synchrotron Radiation Research Center, 
Japan Atomic Energy Agency, 
Sayo, Hyogo 679-5148, Japan} 

\author{S.-i.~Fujimori}
\affiliation{Synchrotron Radiation Research Center, 
Japan Atomic Energy Agency, 
Sayo, Hyogo 679-5148, Japan} 

\author{T.~Okane}
\affiliation{Synchrotron Radiation Research Center, 
Japan Atomic Energy Agency, 
Sayo, Hyogo 679-5148, Japan} 

\author{K.~Terai}
\affiliation{Synchrotron Radiation Research Center, 
Japan Atomic Energy Agency, 
Sayo, Hyogo 679-5148, Japan} 

\author{Y.~Saitoh}
\affiliation{Synchrotron Radiation Research Center, 
Japan Atomic Energy Agency, 
Sayo, Hyogo 679-5148, Japan} 

\author{Y.~Muramatsu}
\affiliation{Synchrotron Radiation Research Center, 
Japan Atomic Energy Agency, 
Sayo, Hyogo 679-5148, Japan}

%\author{T.~Mizokawa}%
%\affiliation{Department of Complexity Science and Engineering, 
%University of Tokyo, Kashiwa, Chiba 277-8561, Japan
%}%

\author{A.~Fujimori}%
\affiliation{Department of Physics, 
University of Tokyo, Bunkyo-ku, Tokyo 113-0033, Japan
}%
\affiliation{Synchrotron Radiation Research Center, 
Japan Atomic Energy Agency, 
Sayo, Hyogo 679-5148, Japan} 

\author{A.~Tanaka}%
\affiliation{Graduate School of Advanced Sciences of Matter, 
Hiroshima University, Higashi-Hiroshima, Hiroshima 739-8530, Japan
}%

\author{H.~Saito}%
\affiliation{Nanoelectronics Research Institute, National 
Institute of Advanced Industrial Science and Technology (AIST), 
Tsukuba Central 2, Umezono 1-1-1, Tsukuba, Ibaraki 305-8568, Japan
}%

\author{K.~Ando}%
\affiliation{Nanoelectronics Research Institute, National 
Institute of Advanced Industrial Science and Technology (AIST), 
Tsukuba Central 2, Umezono 1-1-1, Tsukuba, Ibaraki 305-8568, Japan
}%

\date{\today}% It is always \today, today,

\begin{abstract}
We have performed x-ray magnetic circular dichroism (XMCD) 
and valence-band photoemission studies of the diluted 
ferromagnetic semiconductor Zn$_{1-x}$Cr$_x$Te. 
XMCD signals due to ferromagnetism 
were observed at the Cr 2$p$ absorption edge. 
Comparison with atomic multiplet calculations suggests that the 
magnetically active component of the Cr ion 
was divalent under the tetrahedral crystal field 
with tetragonal distortion along the crystalline $a$-, $b$-, and $c$-axes. 
In the valence-band spectra, spectral weight near the Fermi level was 
strongly suppressed, suggesting 
the importance of Jahn-Teller effect and the strong Coulomb 
interaction between the Cr 3$d$ electrons. 
\end{abstract}

\maketitle

Diluted magnetic semiconductors (DMSs) \cite{Furdyna} showing high 
ferromagnetic Curie temperatures ($T_{\rm C}$'s) 
are considered to be key materials for 
spintronics \cite{DasSarma}. Ferromagnetism of 
II-VI-semiconductor-based DMS Zn$_{1-x}$Cr$_x$Te thin films 
reported by Saito {\it et al.}\ \cite{SaitoPRL, SaitoJAP03} 
has attracted much interest since the $T_{\rm C}$ was 
as high as 300 K and the large $s$$p$-$d$ exchange constant, $N\beta$, 
was confirmed by 
magnetic circular dichroism (MCD) measurements 
in the visible to ultraviolet regions \cite{Ando_Science}. 
Subsequently, it was reported that control of ferromagnetism is 
possible through co-doping N or I during the thin film growth 
\cite{Ozaki_APL, Ozaki_PRL, Kuroda}. 
The ferromagnetism of Zn$_{1-x}$Cr$_x$Te is qualitatively different from 
those of well-known III-V-semiconductor-based ferromagnetic DMSs such as 
Ga$_{1-x}$Mn$_x$As and In$_{1-x}$Mn$_x$As: 
(1) The carrier concentration of Zn$_{1-x}$Cr$_x$Te 
is as low as 
$\sim$10$^{15}$ cm$^{-3}$ and the transport is 
semiconducting \cite{SaitoJAP02, SaitoPRL}, 
whereas ferromagnetism in Ga$_{1-x}$Mn$_x$As 
and In$_{1-x}$Mn$_x$As 
appears for high hole concentration ($\sim$10$^{18}$-10$^{20}$\,cm$^{-3}$) 
and is most likely carrier induced 
\cite{OhnoJMMM}; 
(2) $N\beta$ of Zn$_{1-x}$Cr$_x$Te is positive, that is, 
the hole spin created in the valence band tend to align 
in the same direction as the 
Cr 3$d$ local spin, 
whereas those of Ga$_{1-x}$Mn$_x$As 
and In$_{1-x}$Mn$_x$As are negative \cite{OhnoJMMM, Ando}. 
There have been several studies 
to understand the positive $N\beta$ in Zn$_{1-x}$Cr$_x$Te 
from the electronic structure point of view \cite{MacPRL, MacPRB, 
Bhattacharjee92, Blinowski92, Bhattacharjee94, Mizokawa}. 
In this paper, we report on soft-x-ray MCD (XMCD) in Cr 2$p$ 
core-level absorption and valence-band photoemission (PES) 
measurements to provide information about the 
electronic structure of ferromagnetic Zn$_{1-x}$Cr$_x$Te 
thin films. 
Detailed information such as the valence and crystal field of 
the magnetically active Cr site has been obtained since 
XMCD is an element-specific probe which is sensitive only 
to magnetically active component. 
%\cite{Ohldag, Ueda, Keavney, Edmonds, Rader, Masaki, Mamiya, Ishida}. 

A 150-nm thick epitaxial thin film of 
Zn$_{1-x}$Cr$_x$Te with $x$=0.045 ($T_{\rm C}$$\sim$70\,K) was grown on 
a 120-nm thick ZnTe layer on a 20-nm GaAs buffer layers 
prepared on a semi-insulating GaAs(001) 
substrate using the molecular beam epitaxy method as described elsewhere 
\cite{SaitoPRL}. 
The sample surface was capped with a 3-nm thick ZnTe layer to avoid 
contamination for the Zn$_{1-x}$Cr$_x$Te layer. 
Core-level absorption (XAS) spectra were recorded in the 
total-electron-yield mode at undulator beam line BL23SU 
of SPring-8 \cite{Okamoto_23SU}. The degree of circular polarization 
was higher than 90 \%. The monochromator 
resolution was $E$$/$$\varDelta$$E$$>$10000.
Using a superconducting magnet, magnetic fields 
$H$ up to 7\,T were applied parallel and antiparallel to the propagation 
vector of the incident light and the sample surface. 
Photon helicity was switched 
at each photon energy. 
Valence-band PES measurements 
were performed at BL-18A of Photon Factory (PF), 
High Energy Accelerator Research Organization. 
An $x$=0.043 sample ($T_{\rm C}$$\sim$70\,K) without a 
capping layer and a ZnTe film as a reference were 
measured. The surface was cleaned by Ar-ion sputtering 
at 1.0 kV and subsequent annealing at 200$^{\circ}$C. 
Cleanliness of the sample surface was checked 
by the absence of O 1$s$ core-level PES signal. 
In order to avoid charging effects, the spectra were 
taken at 
room temperature and at $\sim$450\,K. 
The base pressure was $\gtrsim$7.5$\times$10$^{-10}$\,Torr, and the 
resolution of the spectrometer (VG CLAM) including temperature broadening 
was $\sim$200\,meV. 

\begin{figure}[htb]
\begin{center}
\includegraphics[width=8.2cm]{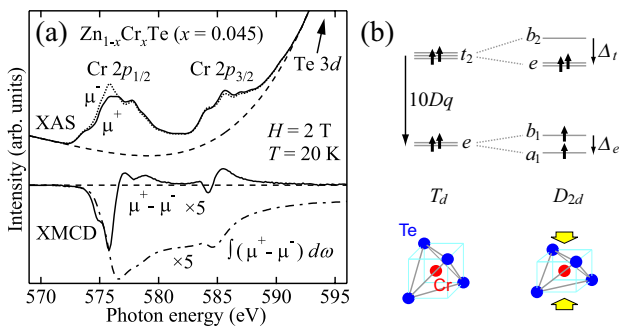}
\caption{\label{fig1}
Cr 2$p$ XAS and XMCD spectra (a) Experimental 
XAS and XMCD spectra. The dashed curve is 
a polynomial fit to the tail of the broad Te 3$d$ absorption. 
The dot-dashed curve shows the energy integral 
of the XMCD. 
(b) Single-electron energy levels in a 
$T_d$ (left) and in a $D_{2d}$ symmetry (right). 
The $t_2$ ($e$) states under the $T_d$ symmetry 
are split into $b_2$ and $e$ ($b_1$ and $a_1$) 
under the $D_{2d}$ symmetry.}
\end{center}
\end{figure}

Figure \ref{fig1}(a) shows the XAS and XMCD spectra taken at the 
Cr 2$p$ absorption edge 
at $T$=20\,K and $H$=2\,T. 
Here, $\mu^+$ and $\mu^-$ indicate absorption spectra for 
photon helicity parallel and antiparallel to the 
Cr 3$d$ spin, respectively. The structures 
in the XAS spectra around $h\nu$=576 and 586 eV are due to absorption from 
the Cr 2$p$$_{3/2}$ and 
Cr 2$p$$_{1/2}$ core levels, respectively. The Cr 2$p$ absorption overlaps 
with the tail of the broad absorption due to Te 3$d$$\to$Te 5$p$ transition. 
Since the Te 5$p$ band is broad and structureless, 
we assumed a polynomial backgroud to separate the 
Te 3$d$ absorption from Cr 2$p$ absorption, 
as shown by dashed curves in Fig.\,\ref{fig1}(a). 
The XAS and XMCD spectra at the Cr 2$p$ edge show multiplet 
structures, which are fingerprints of the localized nature of the Cr 3$d$ 
electrons in a crystal field. 

Following the orbital sum rule \cite{Thole}, 
the energy integral of the XMCD signal in the entire region of the Cr 2$p$ 
absorption is proportional to the orbital moment of the Cr 3$d$ 
electrons, $\langle$$L_z$$\rangle$. 
If the Cr ions are isovalently substituting Zn and become Cr$^{2+}$ 
(3$d^4$) under a crystal field of $T_d$ symmetry as shown in the 
left panel of Fig.\,\ref{fig1}(b), $\langle$$L_z$$\rangle$ 
becomes negative or the integrated XMCD becomes positive 
since the filling of the 3$d$ orbital is less than half 
and there is orbital degrees of freedom in the $t_2$ states. 
However, as shown in Fig.\ \ref{fig1}(a), the integral became 
very small or slightly negative, indicating that 
$\langle$$L_z$$\rangle$ is largely quenched 
compared with the value of the Cr$^{2+}$ ion in the $T_d$-symmetry 
crystal field. 
In fact, Vallin and co-workers have shown in their optical absorption 
study that dilute Cr$^{2+}$ ions in 
bulk ZnTe are  subject to Jahn-Teller distortion \cite{Vallin1, Vallin2} 
as shown in the right panel of Fig.\,\ref{fig1}(b), 
leading to the lift of the orbital degeneracy and 
the quenching of $\langle$$L_z$$\rangle$. 

\begin{figure}[htb]
\begin{center}
\includegraphics[width=8.4cm]{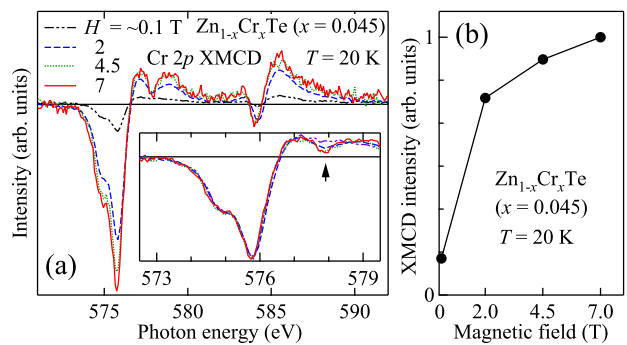}
\caption{\label{fig2}
(Color online) Magnetic-field dependence of XMCD in 
the Cr 2$p$ absorption of 
Zn$_{1-x}$Cr$_{x}$Te ($x$=0.045). (a) XMCD at various $H$ 
taken at $T$=20 K. Inset shows the normalized XMCD. 
(b) Intensity of Cr 2$p$ XMCD 
at $T$=20\,K as a function of $H$.}
\end{center}
\end{figure}

Figure \ref{fig2}(a) shows $H$ dependence of 
XMCD at the Cr 2$p$ absorption edge. The strength of XMCD 
increased with increasing $H$. 
We could observe dichroism down to 
$H$$\sim$0.1\,T as shown in panels (a) and (b), 
indicating the presence of substantial residual magnetization 
of the Cr 3$d$ electrons. The convex behavior of the XMCD strength 
above $H$=2\,T [panel (b)] 
suggests that there were 
Cr ions showing superparamagnetism \cite{Kuroda}
as well as paramagnetism 
in the present sample. 
In the inset of Fig.\ \ref{fig2}(a), 
we show XMCD spectra around the 
Cr 2$p$$_{3/2}$ peak normalized to the XMCD peak intensity. 
The normalized XMCD spectra overlap with each other 
except for a dip structure around $h\nu$$\sim$578 eV 
which is discussed later. 
This indicates that there were a single chemical environment for the 
magnetically active Cr ions in Zn$_{1-x}$Cr$_x$Te. 
This is in line with the unchanged line shapes of the 
visible to ultraviolet MCD under 
varying magnetic field \cite{SaitoPRB, SaitoPRL}. 

\begin{figure}[htb]
\begin{center}
\includegraphics[width=8cm]{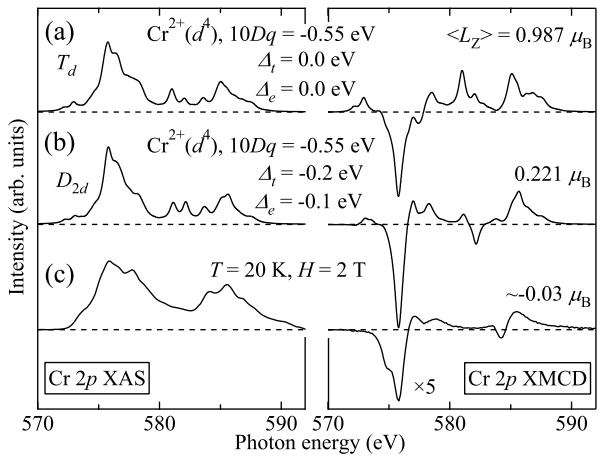}
\caption{\label{fig3}
Cr 2$p$ XAS and XMCD spectra calculated using 
atomic multiplet theory for Cr$^{2+}$ ($d^4$) 
in the $T_d$-symmetry crystal field (a), 
in the $D_{2d}$-symmetry crystal field (b; see, text), 
and the experimental spectra (c). 
Background-subtracted ($\mu^++\mu^-$)/2 has been 
adopted for the experimental XAS spectrum. 
The orbital moment per Cr ion in the ground 
state is also indicated for each panel. }
\end{center}
\end{figure}

In Fig.\ \ref{fig3}, we compare the XAS and XMCD spectra with those 
calculated using atomic multiplet theory taking into account the 
Jahn-Teller distortion. 
The calculated spectra for $T_d$ and $D_{2d}$ symmetries with 
high-spin configurations are shown in Fig.\,\ref{fig3}(b) and (c), 
respectively. 
The crystal-field parameters for $D_{2d}$ symmetry, namely, 
10$D$$q$=-0.55 eV and $\varDelta_t$=-0.20 eV, and $\varDelta_e$=-0.10 eV 
have been adopted from Ref.\ \cite{Vallin1}. 
In the calculation of the spectra under $D_{2d}$ symmetry, we have assumed 
that there are equal numbers of tetragonally-distorted Cr sites 
along the $a$, $b$ and $c$ axes because the film was relaxed 
because of the large mismatch in the 
lattice constant ($a$=6.10\,\AA for ZnTe and $a$=5.65\,\AA for GaAs) 
and free from the epitaxial strain from the substrate \cite{SaitoPRB}. 
In Fig.\ \ref{fig3}, we have also 
indicated $\langle$$L_z$$\rangle$. 
The experimental $\langle$$L_z$$\rangle$ has been 
calculated using the orbital sum rule \cite{Thole}. One can see the 
suppressed $\langle$$L_z$$\rangle$ in the case of $D_{2d}$ compared to the 
case of $T_d$, which is 
in accord with the observed suppression of the Cr 3$d$ orbital moment 
(Fig.\,\ref{fig1}). 
Furthermore, the XMCD lineshape is well reproduced in the calculated 
spectra for the $D_{2d}$ symmetry compared to that for the 
$T_{d}$ symmetry particularly in the Cr 2$p$$_{3/2}$ absorption region. 
The XAS lineshape is also well reproduced for the $D_{2d}$ symmetry 
except for a peak around 578\,eV, which may originate from a small amount 
of other Cr compound(s). In fact, XMCD around 578\,eV showed 
a $H$ dependence different from the other XMCD features, 
as described above. 
Thus, we conclude that the experimental spectra are well reproduced 
if the Jahn-Teller effect is included. 
The lift of the orbital degeneracy due to Jahn-Teller effect 
will directly affect the $s$$p$-$d$ exchange interaction 
as studied theoretically in Refs.\,\cite{Bhattacharjee, Mac, Mizokawa}.

The observed strength of XMCD at the Cr 2$p$ absorption 
edge was $\sim$20 \% ($\sim$0.4 $\mu_{\rm B}$ per Cr ion) of the 
theoretical value where the spin is fully polarized [Note that, 
in Fig.\ \ref{fig3}(b) and (c), the 
experimental dichroism is multiplied by a factor of 
5.] This value is somewhat smaller than 
that derived from the magnetization measurements, for example, 
2.2$\pm$0.3 $\mu_{\rm B}$ reported for $x$=0.035 sample \cite{SaitoPRB}. 
It is possible that 
there is a magnetically dead layer near the surface region, and some 
portion of the Cr ion contributes either to paramagnetism 
or to antiferromagnetism. 

\begin{figure}[htb]
\begin{center}
\includegraphics[width=5.0cm]{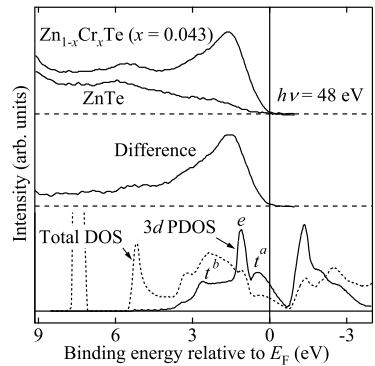}
\caption{\label{fig4}
Valence-band PES spectra of Zn$_{1-x}$Cr$_{x}$Te 
($x$=0.043) and ZnTe and their difference spectra. 
The theoretical total density of states (DOS) and Cr 3$d$ partial 
DOS (PDOS) are taken from Ref.\ \cite{LDA}.} 
\end{center}
\end{figure}

Finally, we show the valence-band PES spectra of 
Zn$_{1-x}$Cr$_x$Te ($x$=0.043) and ZnTe thin films in Fig.\ \ref{fig4}. 
At the excitation energy of $h\nu$=48\,eV, 
the atomic photoionization cross-section of 
Cr 3$d$ is $\sim$20 times larger than that of 
Te 5$p$ \cite{Lindau}. 
In fact, the intensity of the valence band of Cr-doped ZnTe is 
$\sim$2 times higher than the valence-band intensity of 
ZnTe mainly composed of Te 5$p$ states as shown in the top panel of 
Fig.\ \ref{fig4}. 
By subtracting the spectrum of ZnTe from that of Zn$_{1-x}$Cr$_x$Te 
($x$=0.043), we have deduced the Cr 3$d$ partial density 
of states (PDOS) as shown in the middle panel of Fig.\ \ref{fig4}. 
For comparison, we show the calculated total 
density of states (DOS) and the Cr 3$d$ PDOS 
of Zn$_{1-x}$Cr$_x$Te ($x$=0.25) calculated within the 
local-spin-density-function approximation \cite{LDA} 
in the bottom panel of Fig.\ \ref{fig4}. 
One can see a peak at $\sim$1.5 eV and a broad shoulder centered at 
$\sim$3 eV in the experimental Cr 3$d$ PDOS, 
which can be assigned to nonbonding Cr $e$ and 
bonding Cr $t_2$ states, 
and thus agreement between experiment and theory is good on the 
high binding energy side of the valence band. 
On the other hand, the spectral weight near the Fermi level was strongly 
suppressed in the experimental spectra compared to the theoretical 
Cr 3$d$ PDOS which shows half-metallic 
behavior \cite{LDA}. 
This discrepancy may be due to the small Cr concentration of 
$x$=0.043 in the present study, which was not enough for an impurity 
band to form, or because the experiment was performed at room temperature 
in the paramagnetic phase. 
Since the transport properties 
of Zn$_{1-x}$Cr$_x$Te are $p$-type semiconducting 
\cite{SaitoJAP02} even at high 
doping levels of $x$$\sim$0.20 \cite{SaitoPRL}, 
we believe that the suppression of the spectral 
weight at the Fermi level is common to the ferromagnetic 
Zn$_{1-x}$Cr$_{x}$Te. 
If the tetragonal distortion splits the 
Cr 3$d$ $t_2$ orbitals in the $T_d$ symmetry 
as suggested from the XMCD study, 
a gap will open in the Cr 3$d$ $t_2$ impurity band 
which may explain the suppressed spectral weight 
around the Fermi level. A strong Coulomb interaction 
between Cr 3$d$ electrons will also suppress the 
spectral weight at the Fermi level. 
Therefore, it is important to include the distortion effect 
as well as the strong Coulomb interaction effect to understand 
the ferromagnetic Zn$_{1-x}$Cr$_x$Te. 

In summary, we have confirmed the Cr 3$d$ origin of the 
ferromagnetism in Zn$_{1-x}$Cr$_{x}$Te ($x$=0.045) thin film by 
XMCD measurements. Comparisom with atomic multiplet calculation 
idicated that the Cr ion is divalent and substituting the 
Zn site of the host ZnTe and that the Cr ions were subject to 
tetragonal distortion due to Jahn-Teller effect 
as observed in bulk Zn$_{1-x}$Cr$_x$Te \cite{Vallin1, Vallin2}. 
The suppressed spectral weight near the Fermi level in the 
valence-band spectra was discussed in terms of the distortion effect and 
the strong Coulomb effect between Cr 3$d$ electrons. 

We thank T.~Okuda, A.~Harasawa, and T.~Kinoshita for help during 
the experiment at PF, and T.~Mizokawa for useful discussion and comments. 
This work was supported by a Grant-in-Aid for Scientific Research 
in Priority Area ``Semiconductor Nanospintronics" (14076209)
from MEXT, Japan. 
The experiment at PF was approved by the Photon 
Factory Program Advisory Committee (Proposal No.\ 04G002).


\begin{thebibliography}{10}

\bibitem{Furdyna}
J.~K. Furdyna, J. Appl. Phys. {\bf 64},  R29  (1988).

\bibitem{DasSarma}
I. \v{Z}uti{\' c}, J. Fabian, and S.~D. Sarma, Rev. Mod. Phys. {\bf 76},  323
  (2004).

\bibitem{SaitoPRL}
H. Saito, V. Zayets, S. Yamagata, and K. Ando, Phys. Rev. Lett. {\bf 90},
  207202  (2003).

\bibitem{SaitoJAP03}
H. Saito, V. Zayets, S. Yamagata, and K. Ando, J. Appl. Phys. {\bf 93},  6796
  (2003).

\bibitem{Ando_Science}
K. Ando, Science {\bf 312},  1883  (2006).

\bibitem{Ozaki_APL}
N. Ozaki, I. Okabayashi, T. Kumekawa, N. Nishizawa, S. Marcet, S. Kuroda, and
  K. Takita, Appl. Phys. Lett {\bf 87},  192116  (2005).

\bibitem{Ozaki_PRL}
N. Ozaki, I. Okabayashi, T. Kumekawa, N. Nishizawa, S. Marcet, S. Kuroda, and
  K. Takita, Phys. Rev. Lett. {\bf 97},  037201  (2006).

\bibitem{Kuroda}
S. Kuroda, N. Nishizawa, K. Takita, M. Mitome, Y. Bando, K. Osuch, and T.
  Dietl, Nature Mat. {\bf 6},  440  (2007).

\bibitem{SaitoJAP02}
H. Saito, W. Zayets, S. Yamagata, and K. Ando, J. Appl. Phys. {\bf 91},  8085
  (2002).

\bibitem{OhnoJMMM}
H. Ohno, J. Magn. Magn. Mat. {\bf 200},  110  (1999).

\bibitem{Ando}
K. Ando, in {\it Magneto-Optics} {\bf 69},  edited by S. Sugano and N. Kojima,
  Springer Series in Solid  (Springer, Berlin, 2000).

\bibitem{MacPRL}
W. Mac, N.~T. Khoi, A. Twardowski, J.~A. Gaj, and M. Demianiuk, Phys. Rev.
  Lett. {\bf 71},  2327  (1993).

\bibitem{MacPRB}
W. Mac, A. Twardowski, and M. Demianiuk, Phys. Rev. B {\bf 54},  5528  (1996).

\bibitem{Bhattacharjee92}
A.~K. Bhattacharjee, Phys. Rev. B {\bf 46},  5266  (1992).

\bibitem{Blinowski92}
J. Blinowski and P. Kacman, Phys. Rev. B {\bf 46},  12298  (1992).

\bibitem{Bhattacharjee94}
A.~K. Bhattacharjee, Phys. Rev. B {\bf 49},  13987  (1994).

\bibitem{Mizokawa}
T. Mizokawa and A. Fujimori, Phys. Rev. B {\bf 56},  6669  (1997).

\bibitem{Okamoto_23SU}
J. Okamoto, K. Mamiya, S.-I. Fujimori, T. Okane, Y. Saitoh, Y. Muramatsu, A.
  Fujimori, S. Ishiwata, and M. Takano, AIP Conf.\ Proc.\ {\bf 705},  1110
  (2004).

\bibitem{Thole}
B.~T. Thole, P. Carra, F. Sette, and G. van~der Laan, Phys. Rev. Lett. {\bf
  68},  1943  (1992).

\bibitem{Vallin1}
J.~T. Vallin, G.~A. Slack, S. Roberts, and A.~E. Hughes, Phys. Rev. B {\bf 2},
  4313  (1970).

\bibitem{Vallin2}
J.~T. Vallin and G.~D. Watkins, Phys. Rev. B {\bf 9},  2051  (1974).

\bibitem{SaitoPRB}
H. Saito, V. Zayets, S. Yamagata, and K. Ando, Phys. Rev. B {\bf 66},  081201R
  (2002).

\bibitem{Bhattacharjee}
A.~K. Bhattacharjee, Phys. Rev. B {\bf 49},  13987  (1994).

\bibitem{Mac}
W. Mac, A. Twardowski, and M. Demianiuk, Phys. Rev. B {\bf 54},  5528  (1996).

\bibitem{LDA}
T. Fukushima, K. Sato, H. Yoshida, and P.~H. Dederichs, Jpn.\ J.\ Appl.\ Phys.\
   (2004).

\bibitem{Lindau}
I. Lindau, Atomic Data Nucl. Data Tables {\bf 32},  1  (1985).

\end{thebibliography}
\end{document}